# Analysis of shot noise in the detection of ultrashort optical pulse trains

Franklyn Quinlan[*], Tara M. Fortier, Haifeng Jiang, Scott A. Diddams
National Institute of Standards and Technology
325 Broadway, Boulder, CO 80305, USA
*Corresponding author: fquinlan@boulder.nist.gov

**Abstract**: We present a frequency domain model of shot noise in the photodetection of ultrashort optical pulse trains using a time-varying analysis. Shot noise-limited photocurrent power spectral densities, signal-to-noise expressions, and shot noise spectral correlations are derived that explicitly include the finite response of the photodetector. It is shown that the strength of the spectral correlations in the shot noise depends on the optical pulse width, and that these correlations can create orders-of-magnitude imbalance between the shot noise-limited amplitude and phase noise of photonically generated microwave carriers. It is also shown that only by accounting for spectral correlations can shot noise be equated with the fundamental quantum limit in the detection of optical pulse-to-pulse timing jitter.

# 1. Introduction

State-of-the-art photonic applications, such as high-speed signal processing [1], time and frequency dissemination [2, 3], arbitrary optical and rf waveform generation [4-6], and microwave signal generation for coherent radar and microwave atomic clocks [7-11], increasingly rely on stable trains of ultrashort optical pulses, particularly from mode-locked lasers. These applications nearly always involve high-speed photodetection, either to produce the required microwave signals or to assess the quality of the optical waveform. In either case, shot noise presents a fundamental limit on the fidelity of the photodetected signal. Until recently, the relatively low power handling of high-speed photodetectors has kept the signal-to-noise at or near the limit imposed by electronic noise sources. Improvements in the power handling capability of high-speed photodetectors have now increased the achievable microwave power such that the shot noise level is significantly higher than the electronic noise over a broad frequency range [12]. With shot noise-limited detection of high-speed optical waveforms now attainable, correct analysis of the impact of shot noise on measurement fidelity becomes essential.

Most treatments of photocurrent shot noise assume constant intensity on the detector, or, more precisely, that the statistics of the optical intensity are stationary [13-15]. In the absence of quantum optical squeezing, this leads to the well-known shot noise current variance of $\sigma_i^2 = 2qI_{avg}\Delta f$, where $q$ is the fundamental charge, $I_{avg}$ is the average photocurrent, and $\Delta f$ is the measurement bandwidth. Yet the assumption of stationarity is clearly violated in the detection of signals with time-varying intensity, such as ultrashort optical pulses. A more general description of the shot noise current variance that includes time-varying signals is encapsulated in Campbell's theorem [16-18]. This more general formulation was used by Ref. [19], which pointed out that for nonstationary photon rates, only through time averaging can one directly link the shot noise current variance to the average photocurrent. In Refs. [20-23], a frequency domain description of the shot noise of time-varying signals revealed that spectral correlations explained a reduction in the sensitivity of some gravitational wave detectors. Time-varying shot noise models for vacuum tubes and junction diodes have also been developed [24-27], where again there are deviations from the standard, stationary shot noise model.

Despite the extensive work on nonstationary shot noise, to our knowledge all analytical developments of the impact of shot noise on the detection of ultrashort optical pulses have assumed stationary statistics. Recently we experimentally demonstrated optical pulse timing measurements exhibiting significant deviations from the standard, stationary shot noise assumptions [28]. This was explained heuristically in terms of correlations in the shot noise spectrum that result from a frequency comb heterodyning against vacuum fluctuations.

Here we complement our experimental results with a semi-classical, frequency domain description of the shot noise-limited photocurrent of optical signals with time-varying intensity. The focus is on the photodetection of a periodic train of ultrashort optical pulses, which produces a train of electrical pulses (see Fig. 1). In the frequency domain the electrical pulse train corresponds to an array of photonically generated microwave carriers at the pulse repetition rate and its harmonics. This paper addresses the impact of shot noise on measurements of these microwave carriers. A summary of the major results is as follows:

1. Frequency domain expressions of the shot noise-limited signal-to-noise ratio (SNR) are presented for photonically generated microwave signals. In general, the finite detector response must be taken into account to correctly predict the measured SNR. However, a simplified SNR expression for ultrashort pulse detection (Eq. 13) shows one may ignore the detector response in this case. We also predict that, in some cases, the shot noise-limited SNR of a photonically generated microwave signal is not adversely impacted when the microwave power saturates.

2. Expressions for spectral correlations of the photocurrent shot noise are presented. As no correlations are present under CW illumination, they are the heart of the difference between the detection of CW light and signals with periodically-varying intensity. It is shown that, in the shot noise limit, the upper and lower sidebands of photonically generated microwave signals are correlated, and the magnitude of the correlation depends on the optical pulse width (Eq. 17).

3. Expressions for the shot noise-limited amplitude noise and phase noise of photonically generated microwave signals are derived. Due to correlations in the shot noise spectrum, the shot noise-limited phase noise can be orders-of-magnitude below the amplitude noise, depending on the width of the optical pulse (Eq. 20). Also, for certain optical pulse shapes and durations, the shot noise-limited microwave phase noise may exceed the amplitude noise. The pulse shape dependence of the amplitude and phase noise is explored with the specific examples of Gaussian and square-shaped optical pulse intensity profiles. Simplified expressions in the long pulse (Eq. 23) and short pulse limits of the shot noise-limited amplitude noise (Eq. 24) and phase noise (Eq. 25) are also given.

4. For short optical pulses, the shot noise-limited photocurrent gives the pulse-to-pulse timing jitter of the optical train perturbed by fundamental quantum fluctuations (Eq. 27). In a semi-classical description, pulse-to-pulse quantum fluctuations may be thought of as random variations in both the number of photons per pulse and the photon distribution within a pulse. The randomness in the intrapulse photon distribution results in small deviations in the time of arrival of the pulse, or timing jitter [29]. Here we show that only though a time-varying shot noise analysis can the link be made between this quantum-limited timing jitter and the photocurrent shot noise.

The derivation of these results is given in the following section. It is important to note that the presented photocurrent analysis does not include photodetector nonlinearities that become significant when the energy per optical pulse is large. The photodetector nonlinearities most commonly encountered that could affect shot noise-limited measurements are saturation of the microwave power and amplitude-to-phase (AM-PM) conversion. Phenomenological arguments are made to extend our analysis to include the effects of microwave power saturation. Recent studies have demonstrated techniques to strongly mitigate AM-PM conversion [30, 31], and its impact on the presented analysis is discussed in Section 3. At still higher optical powers, optical nonlinearities, such as two-photon absorption, may be present, but are not considered here. Also not explicitly considered are detectors with gain, such as avalanche detectors and photomultiplier tubes. The relationship between our analysis and quantum optical quadrature squeezing, a generalization to other noise sources, and practical limitations on shot noise-limited measurements are also discussed in Section 3. In Section 4 we conclude.

## 2. Semi-classical photocurrent shot noise analysis

### 2.1 Defining the photocurrent power spectral density

We assume detection of an optical field in the coherent state, i.e. we are not examining the detection of squeezed states [32]. We may therefore apply a semi-classical treatment of the photocurrent to derive the photocurrent spectrum, correlations, and amplitude and phase noise of photonically generated microwave signals. The model we use follows that of Ref. [13], where the photocurrent $i(t)$ is the sum of elementary impulses, given by

$$i(t) = \sum_k X_k h(t - k\Delta t). \tag{1}$$

Here, $h(t)$ is the impulse response of the photodetector, and $X$ is a random variable equal to either 1 or 0, depending on whether a photon is detected in (very short) time interval $\Delta t$. Note that the charge of an electron $q$ is included in $h(t)$ such that

$$\int h(t)dt = q. \tag{2}$$

The probability $p$ of a photodetection event in $\Delta t$ is given by the photon rate and the detector quantum efficiency as [33]

$$p = \frac{\eta}{h\nu} P_{opt}(t)\Delta t \equiv \lambda(t)\Delta t, \tag{3}$$

where $P_{opt}(t)$ is the time-varying optical power, $\eta$ is the quantum efficiency of the detector, $h$ is Planck's constant, and $\nu$ is the average optical frequency. The optical intensity profile, $P_{opt}(t)$, is assumed to be a train of pulses, given by

$$P_{opt}(t) = P_0 \sum_n f(t - nT_r), \tag{4}$$

where $f(t)$ is the profile of a single pulse, and $T_r$ is the pulse period. Since we are only concerned with shot noise, we take $P_{opt}(t)$ as otherwise noiseless. Depictions of optical and electrical pulses are shown in Fig. 1(a) and Fig. 1(b), respectively.

Given the probability of a photodetection event, we treat the photocurrent as a stochastic process from which we wish to obtain the spectral density. From the spectral density, we then derive the shot noise-limited SNR of the photonically generated microwave signals. Since any measurement of this SNR will average over many optical pulses, the time-averaged spectral density is used to represent these measurements. We therefore model the double-sided (positive and negative frequencies) photocurrent spectral density as

$$S_i(f) = \lim_{T \to \infty} \frac{1}{T} \langle |\mathcal{F}_T\{i(t)\}|^2 \rangle \ (A^2/Hz), \tag{5}$$

where $\mathcal{F}_T\{i(t)\}$ denotes the finite-time Fourier transform of $i(t)$, defined over time interval $T$ [34], and $\langle \cdot \rangle$ denotes ensemble expectation value.

## 2.2 Shot noise power and signal-to-noise

As detailed in Appendix A, substitution of Eq. (1) into Eq. (5) gives the photocurrent spectral density of

$$S_i(f) = |H(f)|^2 [\lambda_{avg} + S_\lambda(f)], \tag{6}$$

where $\lambda_{avg}$ is the average photoelectron generation rate, $S_\lambda(f)$ is the photoelectron spectral density, and $H(f)$ is the transfer function of the photodetector, given by the Fourier transform of the impulse response. The average photocurrent $I_{avg}$ is given by $q\lambda_{avg}$. The first term in brackets in Eq. (6) is the shot noise, whose single-sided (positive frequencies only) power in a narrow bandwidth $\Delta f$ centered at frequency $f$ is

$$P_{shot}(f) = 2|H_n(f)|^2 q I_{avg} R \Delta f \ (W). \tag{7}$$

Here the photodetector transfer function has been normalized such that $H(f) = qH_n(f)$ (and $H_n(0) = 1$), and $R$ is the terminating load impedance over which the power is measured. Note that the effective impedance must be modified when the detector has an internal termination [35, 36]. The shot noise power of Eq. (7) is the same result one obtains assuming stationary statistics [14, 15, 37], thus verifying that, when averaged over many pulses, the shot noise power spectral density (PSD) in ultrashort optical pulse detection is consistent with standard shot noise analyses .

For a train of optical pulses, $S_\lambda(f)$ is a series of discrete tones at harmonics of the pulse repetition rate. From Eq. (3), $S_\lambda(f)$ is related to the power spectrum of the optical intensity profile, $S_P(f)$, by

$$S_\lambda(f) = \left(\frac{\eta}{h\nu}\right)^2 S_P(f) \tag{8}$$

where $S_P(f)$ is given by

$$S_P(f) = \lim_{T \to \infty} \frac{1}{T} \langle |\mathcal{F}_T\{P_{opt}(t)\}|^2 \rangle, \tag{9}$$

that is, $S_P(f)$ is the PSD of the optical pulse intensity profile. We stress that $S_P(f)$ does not represent the PSD of the optical electric field. The single-sided microwave power of a harmonic at frequency $nf_r$, measured over $\Delta f$, is given by

$$P_\mu(nf_r) = 2q^2 |H_n(nf_r)|^2 S_\lambda(nf_r) R \Delta f. \tag{10}$$

A depiction of the spectrum of the pulse intensity profile and the photocurrent power spectrum are shown in Fig. 1c and Fig. 1d, respectively.

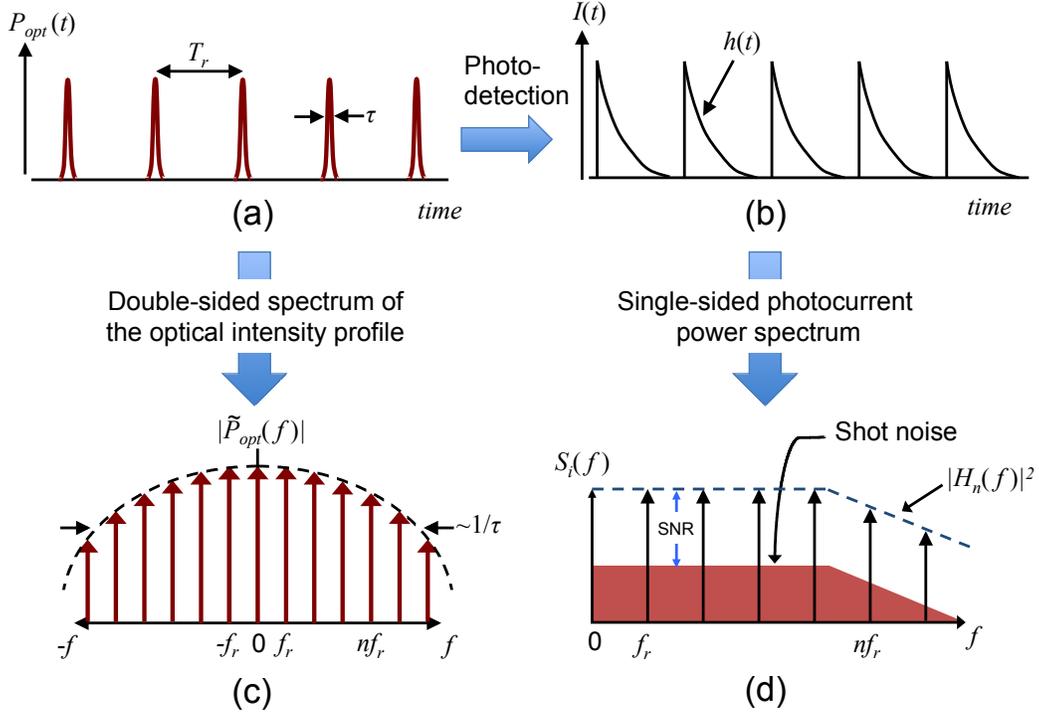

Fig. 1. (Color online) Time and frequency domain depictions of optical and photodetected electrical pulse trains. (a) Optical pulse train intensity profile. (b) Photodetected electrical pulse train when the optical pulse width is much shorter than the photodetector's impulse response time. (c) Spectrum of the optical intensity profile. (d) Power spectrum of the photocurrent.

With the shot noise power and the power of the microwave harmonics now defined, a frequency domain signal-to-noise power ratio (SNR) at the shot noise limit may be expressed as

$$SNR(nf_r) = \frac{P_\mu(nf_r)}{2|H_n(nf_r)|^2 q I_{avg} R \Delta f}. \qquad (11)$$

It is important to note that both the power of the repetition rate harmonic and the shot noise power scale with the photodetector transfer function.

The SNR expression of Eq. (11) can be simplified when the optical pulse width is much less than the detector's impulse response. In this case we may take the double-sided spectral density of the optical intensity profile as a series of delta-functions all of equal magnitude. The single-sided microwave power is then twice the DC power, modified by the detector's transfer function [38]:

$$P_\mu(nf_r) = 2I_{avg}^2 |H_n(nf_r)|^2 R. \qquad (12)$$

Substitution of Eq. (12) into Eq. (11) yields

$$SNR(nf_r) = I_{avg}/(q\Delta f). \qquad (13)$$

The SNR of each microwave carrier increases linearly with average optical power $P_{avg}$ as $I_{avg} = (q\eta/h\nu)P_{avg}$, regardless of the detector's impulse response. Nonlinear behavior of real photodetectors leads to a saturation of microwave power $P_\mu$ when the energy per pulse is high [31, 39, 40]. Although our photocurrent model does not account for photodetector nonlinearity, a phenomenological inclusion to describe the SNR for a photodiode in saturation may be incorporated by considering saturation simply as a power-dependent change in $H_n(f)$ [39-41]. Under this assumption, since the shot noise also scales with $H_n(f)$, Eq. (13) is still valid. A graphical representation of the photocurrent spectrum with a power-dependent $H_n(f)$ is shown in Fig. 2.

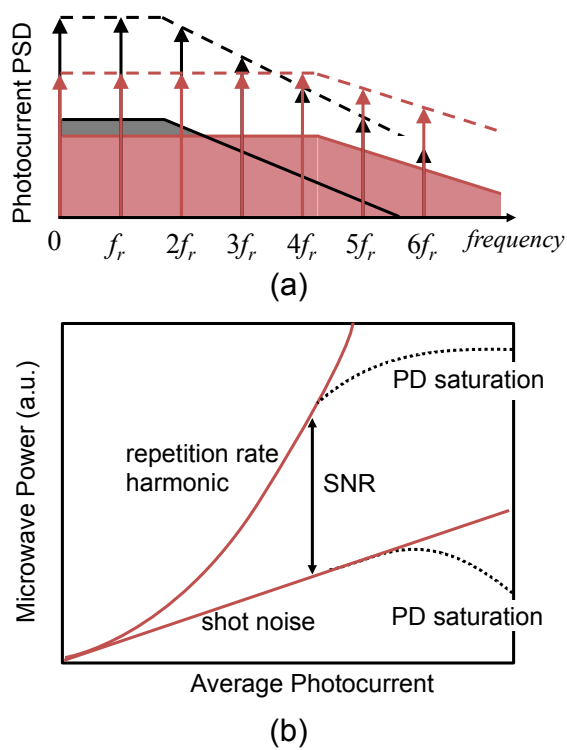

Fig. 2 (Color online) Photodetector saturation modeled as a power-dependent transfer function. (a) Power spectrum of a photodetected train of ultrashort optical pulses under low (red/grey) and high (black) power illumination. Dotted lines represent the roll-off in the response of the photodetector. Shot noise is represented as the shaded regions for low (red/light grey) and high (dark grey) power illumination. In this example, there is a decrease in the photodetector response at high frequencies as the photodetector saturates. This saturation affects the shot noise spectrum as well as the power in the microwave harmonics. (b) The power in one of the microwave harmonics and the shot noise at an adjacent frequency as the average photocurrent is increased. The onset of photodetector saturation leads to a roll-over of the power of the microwave harmonic and a decrease in the shot noise power.

A related quantity is the shot noise-limited relative intensity noise (RIN) spectrum, given by the ratio of the one-sided shot noise PSD to the power at DC [29, 35, 38, 42]:

$$S_{RIN} = \frac{2q}{I_{avg}} = \frac{2h\nu}{\eta P_{avg}} \ (Hz^{-1}). \tag{14}$$

This expression will be compared to the shot noise-limited amplitude noise of a harmonic of the repetition rate in Section 2.4.

## 2.3 Noise correlations

Here we show that there are spectral correlations in the shot noise symmetric about photonically generated microwave signals at harmonics of $f_r$. These correlations manifest themselves, for example, in measurements where the photocurrent is demodulated with a reference signal. Correlated frequency components are then folded upon one another, allowing them to interfere. To determine the correlations in the photocurrent shot noise spectrum, we use the double-sided, time-averaged cross-spectral density of the photocurrent, given by

$$S_i(f, f') = \lim_{T \to \infty} \frac{1}{T} \langle \mathcal{F}_T \left\{ \sum_k X_k h(t - k\Delta t) \right\} \cdot \mathcal{F}_T^* \left\{ \sum_l X_l h(t - l\Delta t) \right\} \rangle. \tag{15}$$

Keeping only the noise term (details are in Appendix A), this reduces to

$$S_{i_n}(f, f') = H(f) H^*(f') \lim_{T \to \infty} \frac{1}{T} [\Lambda_T(f - f')], \tag{16}$$

where $\Lambda_T(f)$ is the finite-time Fourier transform of the photoelectron rate $\lambda(t)$. The noise at frequencies $f$ and $f'$ are said to be correlated if $S_{i_n}(f, f')$ is nonzero. For CW illumination, $S_{i_n}(f, f')$ is nonzero only for $f = f'$, since the frequency content of a photodetected coherent CW signal is restricted to DC. For a train of ultrashort pulses, however, $S_{i_n}(f, f')$ is nonzero whenever $f - f' = nf_r$. Importantly, the upper and lower sidebands symmetric about harmonics of the repetition rate at frequencies $nf_r \pm \delta f$ are also correlated. This may be seen by considering both positive and negative frequencies of the spectrum, as shown in Fig. 3. From Eq. (16), the noise at frequency $nf_r + \delta f$ is correlated with the noise at frequency $-nf_r + \delta f$, with the degree of correlation proportional to $\Lambda_T(2nf_r)$. Since $\lambda(t)$ is real, $\Lambda_T(f) = \Lambda_T^*(-f)$, where "*" denotes complex conjugate. Therefore frequencies $nf_r \pm \delta f$ are correlated. The magnitude of the correlation between frequencies $nf_r \pm \delta f$ is then equal to the magnitude of the correlation between frequencies $nf_r + \delta f$ and $-nf_r + \delta f$, given by [34]

$$|C(nf_r + \delta f, nf_r - \delta f)| = |C(nf_r + \delta f, -nf_r + \delta f)|$$
$$= \frac{|S_{i_n}(nf_r + \delta f, -nf_r + \delta f)|}{[S_i(nf_r + \delta f)S_i(-nf_r + \delta f)]^{1/2}} \qquad (17)$$
$$= \frac{|\tilde{P}_{opt}(2nf_r)|}{\tilde{P}_{opt}(0)}.$$

Here $\tilde{P}_{opt}(f)$ is the Fourier transform of $P_{opt}(t)$, and $|C(nf_r + \delta f, nf_r - \delta f)| \leq 1$. Thus the degree of correlation between sidebands is independent of the photodetector response, depending only on the spectrum of the optical pulse train intensity profile. As illustrated in Fig. 1(c), the ratio of $\tilde{P}_{opt}(2nf_r)$ to $\tilde{P}_{opt}(0)$ scales with the inverse optical pulse width; therefore the shorter the optical pulse, the higher the degree of correlation. The fact that shot noise is correlated between upper and lower sidebands implies that the shot noise does not contribute equally to the phase and amplitude quadratures of the microwave signal. This is shown in the following section.

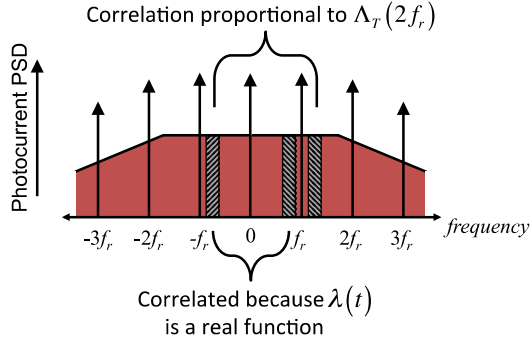

Fig. 3. (Color online) Correlations in the sidebands about a photonically-generated microwave carrier, revealed with a double-sided representation of the photocurrent power spectrum.

## 2.4 Amplitude noise, phase noise and timing jitter

We now consider the shot noise-limited amplitude and phase noise of a photonically generated microwave carrier. From the phase noise we also derive the shot noise-limited pulse-to-pulse timing jitter. A standard method of phase and amplitude noise measurement of a microwave signal is to multiply the microwave carrier with a noiseless reference signal at the same frequency, represented as

$$[1 + a(t)] \cos(2\pi f_0 t + \theta(t)) \cdot \cos(2\pi f_0 t + \Phi_r), \qquad (18)$$

where $a(t)$ and $\theta(t)$ are the zero-mean amplitude and phase fluctuations of the microwave signal under test, respectively, and $\Phi_r$ is the phase offset of the reference signal. When the two signals are in phase ($\Phi_r = 0$) amplitude noise is measured, and when the signals are in quadrature ($\Phi_r = \pi/2$) phase noise is measured [43]. For the detection of ultrashort optical pulses, we model phase and amplitude noise measurements as the isolation of a harmonic of the repetition rate with a bandpass filter followed by multiplication with the reference signal [28]. The demodulated photocurrent is then given by

$$i_m(t) = \left[\left(\sum_k X_k h(t - k\Delta t)\right) * g(t)\right] \cdot \cos(2\pi n f_r t + \Phi_r), \tag{19}$$

where $g(t)$ is the impulse response of a bandpass filter, and $*$ denotes the convolution operation. By deriving the spectral density of the demodulated photocurrent (Appendix B), the single-sideband amplitude or phase noise relative to the signal power at $nf_r$ may be expressed as

$$L_{AM,PM} = \frac{qI_{avg}|H_n(nf_r)|^2 R}{P_\mu(nf_r)}\left[1 \pm \frac{|\tilde{P}_{opt}(2nf_r)|}{\tilde{P}_{opt}(0)}\right. \\ \left. \cdot \cos\left(2\Phi_{opt}(nf_r) - \Phi_{opt}(2nf_r)\right)\right] \;(Hz^{-1}), \tag{20}$$

where the plus sign in the brackets is for amplitude noise and the minus sign is for phase noise, and $\Phi_{opt}(f)$ is the spectral phase of the pulse intensity profile. Note that this is not simply one-half the noise-to-signal ratio of the photocurrent (inverse of Eq. (11)) as is commonly assumed [10, 11, 35, 38, 44-48]. Equation (20) contains the additional term in brackets that is a direct result of shot noise correlations that are symmetric about the microwave carrier, described above in Section 2.3.

Equation (20) is applicable to any photonically generated microwave signal, regardless of the details of the photodetected optical intensity variations. To illustrate the impact of pulse shape and pulse width, we now consider the amplitude and phase noise of microwave signals derived from photodetecting Gaussian and square-shaped optical pulse trains.

### *2.4.1 Gaussian Pulses*

Consider a train of Gaussian-shaped pulses, represented as

$$P_{opt}(t) = \frac{E_p}{\tau_G \sqrt{\pi}} \sum_n \exp\{-((t - nT_r)/\tau_G)^2\}, \tag{21}$$

where $E_p$ is the energy per pulse, $T_r$ is the pulse period, and $\tau_G$ is related to the pulse intensity full-width at half-maximum $\tau_P$ by $\tau_P = 2\sqrt{ln(2)}\tau_G$. By calculating $\tilde{P}_{opt}(f)$ (as shown in Fig. 4a) and substituting into Eq. (20), the amplitude and phase noise of a microwave carrier at frequency $nf_r$ become

$$L_{AM,PM} = \frac{qI_{avg}|H_n(nf_r)|^2 R}{P_\mu(nf_r)}[1 \pm \exp\{-(2\pi n f_r \tau_G)^2\}]. \tag{22}$$

Here we have used the fact that for Gaussian pulses, nonzero $\Phi_{opt}(f)$ is only due to an arbitrary time delay, leading to $2\Phi_{opt}(nf_r) = \Phi_{opt}(2nf_r)$.

It is interesting to examine the noise in the long and short pulse limits. In the long pulse limit the magnitude of the shot noise correlations about $nf_r$ goes to zero, and the term in brackets in Eq. (22) is unity. In this case, the shot noise is equally distributed in amplitude and phase, yielding

$$L_{AM,PM} = \frac{qI_{avg}|H_n(nf_r)|^2 R}{P_\mu(nf_r)}. \tag{23}$$

In the ultrashort pulse limit, Eq. (22) may be simplified by noting $nf_r\tau_G \ll 1$ and using Eq. (12). In this limit

$$L_{AM} = \frac{q}{I_{avg}} = \frac{1}{2}S_{RIN} \tag{24}$$

and

$$L_{PM} = \frac{q}{2I_{avg}}(2\pi nf_r\tau_G)^2. \tag{25}$$

Note that in the short pulse limit, the shot noise resides exclusively in the amplitude quadrature of the microwave signal, and only then can be easily related to the shot noise-limited RIN. The phase noise can be several orders-of-magnitude below the amplitude noise, depending on the optical pulse width. For example, for an average photocurrent of 10 mA, and 100 fs duration optical pulses, the predicted shot noise-limited amplitude and phase noise measured on a 10 GHz harmonic, are $L_{AM} = -168\,\text{dBc/Hz}$, and $L_{PM} = -219\,\text{dBc/Hz}$, respectively, a ratio of ~50 dB. A plot of the amplitude and phase noise deviation from the long pulse limit for Gaussian pulses is shown in Fig. 4(b).

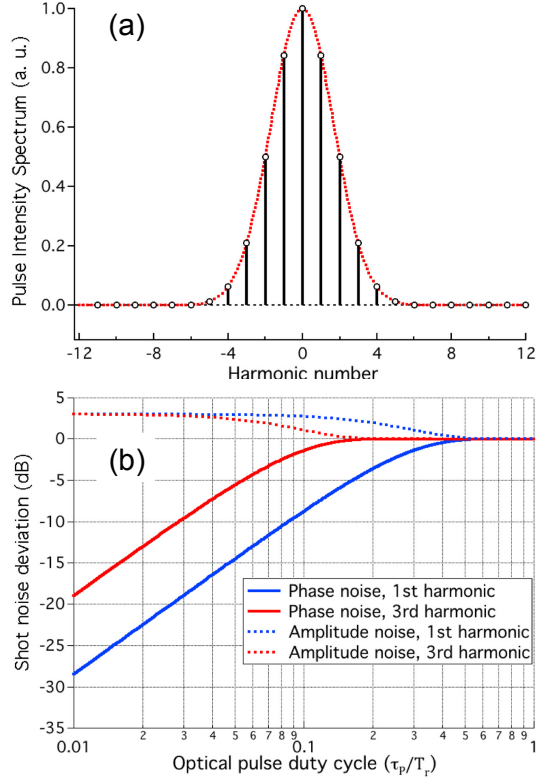

Fig. 4. (Color online) (a) Spectrum of the optical intensity profile of a train of Gaussian-shaped pulses. The pulse duty cycle ($\tau_p/T_r$) is 0.25. (b) Shot noise deviation from the long pulse limit for a train of Gaussian pulses.

The microwave phase noise is often used to determine the optical pulse-to-pulse timing jitter [49]. The shot noise-limited mean-square pulse-to-pulse timing jitter is recovered from the microwave phase noise measured at $nf_r$ by

$$\sigma_t^2 = \frac{1}{(2\pi n f_r)^2} \int_0^{f_r/2} 2L_{PM}(f)df \quad (s^2), \tag{26}$$

where $\sigma_t$ is the r.m.s. timing deviation. Substituting Eq. (25) into Eq. (26) gives

$$\sigma_t^2 = \frac{q}{2I_{avg}} f_r \tau_G^2 = \frac{1}{2\eta} \frac{h\nu}{E_p} \tau_G^2. \tag{27}$$

This differs from the quantum fluctuations-induced variance in the time of arrival of the optical pulses themselves by $1/\eta$ [29]. (The difference arises from the fact that, for $\eta < 1$, there is an additional randomness associated with detection that increases the noise [50].) *Thus only by accounting for correlations in the shot noise spectrum may we conclude that shot noise corresponds to the fundamental quantum limit of the timing precision of ultrashort optical pulses.*

### 2.4.2 Square pulse train

The shot noise behavior for a train of square pulses provides an interesting contrast to the behavior of Gaussian pulses. The pulse intensity profile for a train of square pulses is

$$P_{opt}(t) = \frac{E_p}{\tau_p} \sum_n \Pi\left(\frac{(t - nT_r)}{\tau_p}\right) \tag{28}$$

where

$$\Pi\left(\frac{t}{\tau_p}\right) = \begin{cases} 1 \text{ for } -\tau_p/2 \leq t \leq \tau_p/2 \\ 0 \text{ otherwise} \end{cases}. \tag{29}$$

In this case the shot noise-limited phase and amplitude noise are

$$L_{AM,PM} = \frac{qI_{avg}|H_n(nf_r)|^2 R}{P_\mu(nf_r)} \Big[1 \pm |\text{sinc}(2\pi n\tau_p f_r)| \\ \cdot \cos\left(2\Phi_{opt}(nf_r) - \Phi_{opt}(2nf_r)\right)\Big]. \tag{30}$$

Unlike the Gaussian pulse train, $\Phi_{opt}(f)$ must be included to account for the sign changes in the optical pulse intensity spectrum (shown in Fig. 5(a)) that give an additional phase shift of $\pi$ radians. Just as with Gaussian pulses, the amplitude and phase noise in the long pulse limit are given by Eq (23). For short optical pulses, the amplitude noise is the same as given in Eq. (24), and the phase noise reduces to

$$L_{PM} = \frac{q}{12 I_{avg}} \left(2\pi nf_r\tau_p\right)^2. \tag{31}$$

As with the Gaussian pulse train, calculation of the pulse timing jitter from Eq. (31) gives the quantum fluctuations-induced variance in the time of arrival of square optical pulses. Phase and amplitude noise deviations for a train of square pulses are shown in Fig. 5(b). In between the long and short pulse limits, the amplitude and phase noise deviations oscillate, and the phase noise may exceed the amplitude noise. This is due to the sign changes in the sinc function that, for certain optical pulse widths, gives $2\Phi_{opt}(nf_r) - \Phi_{opt}(2nf_r) = \pi$. However, the error resulting from ignoring this sign change is small, and would only need to be considered if measurement uncertainty within ~1 dB is required.

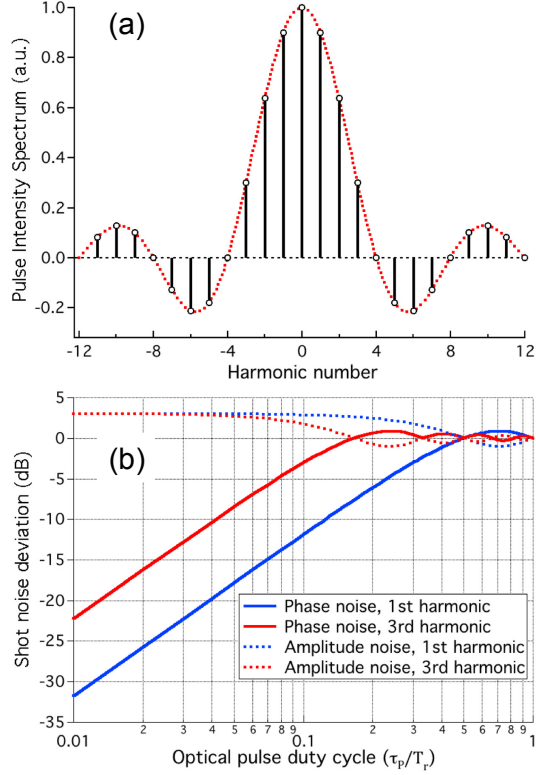

Fig. 5. (Color online) (a) Pulse intensity profile spectrum of a train of square pulses with a duty cycle of 0.25. (b) Shot noise deviation from the long pulse limit for a train of square pulses.

## 3. Discussion

The most significant result of the presented analysis is the predicted shot noise imbalance between amplitude and phase quadratures of photonically generated microwave signals. For the detection of a train of ultrashort optical pulses, the shot noise contribution to the phase noise can be reduced to a negligible level. In the frequency domain, this can be interpreted as correlations in the shot noise spectrum that result in noise cancelation in a phase noise measurement, and constructive interference in an amplitude noise measurement. Recent measurements have confirmed this behavior [28]. The time domain interpretation considers the pulse-to-pulse optical timing jitter as fundamentally limited by the variance in the photon distribution within a pulse, and this jitter is pulse width dependent. Upon photodetection, the photon limit manifests itself as photocurrent shot noise. When the optical pulse-to-pulse jitter is retrieved through a measurement of the phase of a generated microwave carrier, the shot noise shows the same dependence on the optical pulse width as the photon variance.

As mentioned at the beginning of Section 2, the presented analysis is semi-classical and does not apply to the detection of non-classical states, such as squeezed light. Although this work does represent a kind of shot noise

"squeezing," the term squeezing is so closely associated with the detection of non-classical states that we avoid its use here. Quantum optical quadrature squeezing, where quantum-limited noise is shifted from the amplitude to the phase, or vice versa, of the optical field, has an analogous description, however. Quantum optical quadrature squeezing can also be described in terms of spectral correlations, though in this case the noise correlations are in the optical domain [51]. Such correlations change the SNR of the directly detected signal, that is, the shot noise power of Eq. (7) is changed. For a time-variance of the optical intensity, noise correlations appear after the square-law detection of the photodetector. While this results in a phase/amplitude imbalance in the generated microwave field, there is no effect on Eq. (7).

The concept of spectral noise correlations is quite general, and may be considered as the consequence of a cyclostationary noise process [26]. Perhaps it is not surprising then that shot noise is not the only noise source associated with the detection of ultrashort optical pulses that will display a phase/amplitude imbalance on a photonically generated microwave carrier. Any noise on the optical pulse train that is periodically-varying will produce correlations in the photocurrent noise spectrum. It is interesting to note that this should apply to quantum noise in optical amplification. Using, for example, a doped-fiber amplifier, the dominant photocurrent noise resulting from optical amplification is described semi-classically as signal-spontaneous beat noise [52, 53]. When amplifying a train of ultrashort pulses, this signal-spontaneous beat noise should exhibit the same spectral correlation behavior in the photocurrent as shot noise, reducing its impact on the microwave phase. In this case the often ignored spontaneous-spontaneous beat noise may in fact determine the microwave phase noise floor. Correlated optical noise has also been observed in harmonically mode-locked lasers, where the supermode noise power has been shown to be sensitive to the optical pulse width [54]. Another possible time-varying noise source is photodetector flicker noise [45]. If this noise source is present only when current is being generated, then it too can produce phase/amplitude imbalance. However, it is likely that this noise will scale with the electrical, as opposed to the optical, pulse width.

Finally, it is important to consider the practical limitations in measuring large microwave phase/amplitude imbalance in shot noise. The phase noise levels in particular are quite low, and can be well below the thermal noise limit at room temperature. For example, to measure the shot noise-limited phase noise on a 10 GHz harmonic for a 100 fs pulse train generating 10 mA of average photocurrent, the system temperature would need to be < 2 μK. Also, it is possible that AM-PM conversion during photodetection could alter the phase/amplitude shot noise imbalance. Preliminary experimental evidence supports this notion [28]. Although AM-PM conversion has been measured in detectors and quantified in a parameter that converts the measured RIN to the microwave phase noise [30, 31], more work is needed to relate these measurements to a similar conversion in shot noise. Additionally, phase bridge measurements, traditionally used to determine microwave phase noise, can have residual sensitivity to amplitude noise [55]. With the amplitude noise orders-of-magnitude higher than the phase noise, careful attention must be given to reducing the amplitude sensitivity of the phase bridge.

# 4. Conclusion

A complete description of the shot noise spectral density in the detection of ultrashort optical pulses requires incorporating spectral correlations that result from the time-varying nature of the optical intensity. Here we have derived expressions for shot noise correlations, and have shown the impact of these correlations on the phase and amplitude noise of photonically-generated microwave signals. This analysis has allowed us to clearly establish the link between photocurrent shot noise and the fundamental quantum limit of the timing jitter of optical pulses. The analysis reveals a new regime of microwave phase noise sensitivity where the shot noise contribution can be made negligible, provided the optical pulse width on the photodetector is short enough.

# Acknowledgements

We thank P. Winzer, E. Ivanov, J. Davila-Rodriguez, N. Newbury, and N. Ashby for their contributions and comments on this manuscript. This work was supported by NIST and in part by DARPA. It is a contribution of an agency of the US government and is not subject to copyright in the USA.

*Appendix A*

Here we supply details on the derivation of the photocurrent power spectrum and shot noise correlations. To this end we define the photocurrent cross spectrum as

$$S_i(f, f') = \lim_{T \to \infty} \frac{1}{T} \langle \mathcal{F}_T \left\{ \sum_k X_k h(t - k\Delta t) \right\} \times \mathcal{F}_T^* \left\{ \sum_l X_l h(t - l\Delta t) \right\} \rangle, \quad (A1)$$

where the finite-time Fourier transform is defined by

$$\mathcal{F}_T\{x(t)\} \equiv \int_{-T/2}^{T/2} x(t)e^{-i2\pi f t}dt \equiv X_T(f). \tag{A2}$$

Interchanging the summation and Fourier transform operations yields

$$S_i(f,f') = \lim_{T\to\infty}\frac{1}{T}\langle \sum_k X_k \mathcal{F}_T\{h(t-k\Delta t)\} \cdot \sum_l X_l \mathcal{F}_T^*\{h(t-l\Delta t)\}\rangle, \tag{A3}$$

or

$$S_i(f,f') = \lim_{T\to\infty}\frac{1}{T}\sum_k\sum_l \langle X_k X_l\rangle H_T(f)H_T^*(f')\exp\{-i2\pi f k\Delta t\}\exp\{i2\pi f' l\Delta t\}, \tag{A4}$$

where $H_T(f)$ is the finite-time Fourier transform of $h(t)$. Individual photoelectron events are considered statistically independent. The ensemble expectation of the product of random variables $X_k$ and $X_l$ is therefore given by [13]:

$$\langle X_k X_l\rangle = \begin{cases} \lambda(k\Delta t)\Delta t & \text{for } k = l \\ \lambda(k\Delta t)\lambda(l\Delta t)\Delta t\Delta t & \text{for } k \neq l \end{cases}, \tag{A5}$$

where $\lambda(t) = (\eta/h\nu)P(t)$ is the photoelectron generation rate. As $h(t)$ is the impulse response of the detector, we may assume that it is of finite duration, and let $T$ be much greater than the duration of $h(t)$. In this case, $H_T(f) = H(f)$. Using this substitution,

$$S_i(f,f') = H(f)H^*(f')\lim_{T\to\infty}\frac{1}{T}\sum_k \lambda(k\Delta t)\Delta t \exp\{-i2\pi(f-f')k\Delta t\} + \sum_k\sum_{l,l\neq k}(k\Delta t)\lambda(l\Delta t)\Delta t\Delta t \exp\{-i2\pi f k\Delta t\}\exp\{i2\pi f' l\Delta t\}. \tag{A6}$$

Letting $\Delta t \to 0$ yields

$$S_i(f,f') = H(f)H^*(f')\lim_{T\to\infty}\frac{1}{T}\left[\int_{-T/2}^{T/2}\lambda(\tau)e^{-i2\pi(f-f')\tau}d\tau + \iint_{-T/2}^{T/2}\lambda(\tau)\lambda(\tau')e^{-i2\pi f\tau}e^{i2\pi f'\tau'}d\tau d\tau'\right] \tag{A7}$$

or

$$S_i(f,f') = H(f)H^*(f') \lim_{T\to\infty} \frac{1}{T}[\Lambda_T(f-f') + \Lambda_T(f)\Lambda_T^*(f')]. \tag{A8}$$

For $f = f'$, this becomes the photocurrent spectral density:

$$S_i(f) = |H(f)|^2[\lambda_{avg} + S_\lambda(f)] \ (A^2/Hz), \tag{A9}$$

where

$$\lambda_{avg} = \lim_{T\to\infty} \frac{1}{T}\Lambda_T(0) \tag{A10}$$

and

$$S_\lambda(f) = \lim_{T\to\infty} \frac{1}{T}|\Lambda_T(f)|^2. \tag{A11}$$

Correlations in the photocurrent noise may be determined by selecting from Eq. (A8) the noise cross-spectrum:

$$S_{i_n}(f,f') = H(f)H^*(f') \lim_{T\to\infty} \frac{1}{T}[\Lambda_T(f-f')]. \tag{A12}$$

This is the same expression as given in Eq. (16).

### Appendix B

Derivation of the demodulated photocurrent follows similar lines as the development in Appendix A, but is more involved. An abbreviated version of this derivation may be found in Ref. [28]. First, we rewrite Eq. (18) to explicitly include the voltage amplitudes of the signal under test and reference:

$$V_m(t) = V_1[1 + a(t)]\cos(2\pi f_r t + \theta(t)) \cdot V_2\cos(2\pi f_r t + \Phi_r). \tag{B1}$$

For $a(t) \ll 1$ and $\theta(t) \ll 1$ radian, the voltage output of the mixer when $\Phi_r = 0$ is

$$V_m(t) = \frac{V_1 V_2}{2}[1 + a(t)]. \tag{B2}$$

When $\Phi_r = \pi/2$, the mixer output is

$$V_m(t) = \frac{V_1 V_2}{2}\theta(t). \tag{B3}$$

Thus the phase noise is given by the ratio of the in-quadrature condition to the nominal DC value when in-phase. We now turn to the case when the signal under test is the bandpass-filtered photocurrent. In this case we write the spectral density of the mixer output as

$$S_m(f) = \lim_{T\to\infty} \frac{1}{T} \langle \left| \mathcal{F}_T \left\{ \left[ \left( \sum_k X_k h(t - k\Delta t) \right) * g(t) \right] \right. \right.$$
$$\left. \left. \cdot \cos(2\pi f_r t + \Phi_r) \right\} \right|^2 \rangle. \qquad \text{(B4)}$$

Moving the convolution, multiplication and Fourier transform inside the summation yields

$$S_m(f) = \lim_{T\to\infty} \frac{1}{T} \langle \left| \sum_k X_k \mathcal{F}_T\{[h(t-k\Delta t) * g(t)] \right.$$
$$\left. \cdot \cos(2\pi f_r t + \Phi_r)\} \right|^2 \rangle. \qquad \text{(B5)}$$

Evaluating the Fourier transform yields

$$S_m(f) = \lim_{T\to\infty} \frac{1}{T} \langle \left| \sum_k X_k [H_T(f) e^{-i2\pi f k \Delta t} G_T(f)] \right.$$
$$\left. * \left[ \frac{1}{2}\delta(f-f_r) e^{i\Phi_r} + \frac{1}{2}\delta(f+f_r) e^{-i\Phi_r} \right] \right|^2 \rangle, \qquad \text{(B6)}$$

where $G_T(f)$ is the finite-time Fourier transform of $g(t)$, and $\delta(f)$ is the Dirac-delta function. Strictly speaking, the delta function representation requires the evaluation of the limit $T \to \infty$. However, it is a reasonable approximation as long as $1/T \gg f_r$. As in Appendix A, we also let $T$ be much greater than the duration of $h(t)$ and $g(t)$ such that $H_T(f) = H(f)$ and $G_T(f) = G(f)$. Convolution and squaring yields the sum of the following four terms:

$$\frac{1}{4} |H(f-f_r)|^2 |G(f-f_r)|^2$$
$$\cdot \sum_k \sum_l \langle X_k X_l \rangle \qquad \text{(B7a)}$$
$$\cdot \exp\{-i2\pi(f-f_r)k\Delta t\} \exp\{i2\pi(f-f_r)l\Delta t\}$$

$$\frac{1}{4} |H(f+f_r)|^2 |G(f+f_r)|^2$$
$$\cdot \sum_k \sum_l \langle X_k X_l \rangle \qquad \text{(B7b)}$$
$$\cdot \exp\{-i2\pi(f+f_r)k\Delta t\} \exp\{i2\pi(f+f_r)l\Delta t\}$$

$$\frac{1}{4}H(f-f_r)H^*(f+f_r)G(f-f_r)G^*(f+f_r)e^{2i\Phi_r}$$
$$\cdot \sum_k \sum_l \langle X_k X_l \rangle \exp\{-i2\pi(f-f_r)k\Delta t\} \quad \text{(B7c)}$$
$$\cdot \exp\{i2\pi(f+f_r)l\Delta t\}$$

$$\frac{1}{4}H^*(f-f_r)H(f+f_r)G^*(f-f_r)G(f+f_r)e^{-2i\Phi_r}$$
$$\cdot \sum_k \sum_l \langle X_k X_l \rangle \exp\{i2\pi(f-f_r)k\Delta t\} \quad \text{(B7d)}$$
$$\cdot \exp\{-i2\pi(f+f_r)l\Delta t\}$$

Using similar arguments as in Appendix A, expression (B7a) reduces to

$$\frac{1}{4}|H(f-f_r)|^2|G(f-f_r)|^2\left[\lim_{T\to\infty}\frac{1}{T}\Lambda_T(0)+S_\lambda(f-f_r)\right]. \quad \text{(B8)}$$

Likewise, expression (B7b) becomes

$$\frac{1}{4}|H(f+f_r)|^2|G(f+f_r)|^2\left[\lim_{T\to\infty}\frac{1}{T}\Lambda_T(0)+S_\lambda(f+f_r)\right]. \quad \text{(B9)}$$

Expression (B7c) yields

$$\frac{1}{4}H(f-f_r)H^*(f+f_r)G(f-f_r)G^*(f+f_r)e^{2i\Phi_r}$$
$$\cdot\left[\lim_{T\to\infty}\frac{1}{T}\int_{-T/2}^{T/2}\lambda(\tau)e^{i2\pi(2f_r)\tau}d\tau\right.$$
$$\left.+\lim_{T\to\infty}\frac{1}{T}\iint_{-T/2}^{T/2}\lambda(\tau)\lambda(\tau')e^{-i2\pi(f-f_r)\tau}e^{i2\pi(f+f_r)\tau'}d\tau d\tau'\right], \quad \text{(B10)}$$

which may be expressed as

$$\frac{1}{4}H(f-f_r)H^*(f+f_r)G(f-f_r)G^*(f+f_r)e^{2i\Phi_r}\left[\lim_{T\to\infty}\frac{1}{T}\Lambda_T(2f_r)\right.$$
$$\left.+\lim_{T\to\infty}\frac{1}{T}\Lambda_T(f-f_r)\Lambda_T^*(f+f_r)\right]. \quad \text{(B11)}$$

Expression (B7d) reduces to the complex conjugate of expression (B11). To sum these four terms, we exploit the fact that $H(f)$, $G(f)$, and $\Lambda_T(f)$ are the Fourier transforms of real functions, each term is band-limited by $G(f)$, and we only need only to consider the baseband section of each term. Thus we use the following relationship:

$$H(f-f_r)G(f-f_r)=H^*(f+f_r)G^*(f+f_r). \quad \text{(B12)}$$

Note that this relation is equivalent to requiring that the product $H(f)G(f)$ is symmetric about $f_r$. This is necessary for fullest cancellation of the upper and lower sidebands in the phase noise.

Expressions (B8) and (B9) may now be summed to give

$$\frac{1}{2}|H(f-f_r)|^2|G(f-f_r)|^2\left[\lim_{T\to\infty}\frac{1}{T}\Lambda_T(0)+S_\lambda(f-f_r)\right]. \tag{B13}$$

Summation of expression (B11) with its complex conjugate yields

$$\begin{aligned}\frac{1}{2}&|H(f-f_r)|^2|G(f-f_r)|^2 S_\lambda(f-f_r)\\ &\cdot\cos(2\Phi_r+2\Phi_H(f-f_r)+2\Phi_G(f-f_r)+2\Phi_\Lambda(f-f_r))\\ &+\frac{1}{2}|H(f-f_r)|^2|G(f-f_r)|^2\lim_{T\to\infty}\frac{1}{T}|\Lambda_T(2f_r)|\\ &\cdot\cos(2\Phi_r+2\Phi_H(f-f_r)+2\Phi_G(f-f_r)-\Phi_\Lambda(2f_r)).\end{aligned} \tag{B14}$$

The output of the mixer is then given by the summation of (B13) and (B14), or

$$\begin{aligned}S_m(f)=\frac{1}{2}&|H(f-f_r)|^2|G(f-f_r)|^2\left[\lim_{T\to\infty}\frac{1}{T}\Lambda_T(0)+\lim_{T\to\infty}\frac{1}{T}|\Lambda_T(2f_r)|\right.\\ &\left.\cdot\cos(2\Phi_r+2\Phi_H(f-f_r)+2\Phi_G(f-f_r)-\Phi_\Lambda(2f_r))\right]\\ &+\frac{1}{2}|H(f-f_r)|^2|G(f-f_r)|^2 S_\lambda(f-f_r)[1\\ &+\cos(2\Phi_r+2\Phi_H(f-f_r)+2\Phi_G(f-f_r)+2\Phi_\Lambda(f-f_r))].\end{aligned} \tag{B15}$$

For phase noise, $2\Phi_r+2\Phi_H(f-f_r)+2\Phi_G(f-f_r)+2\Phi_\Lambda(f-f_r)=\pi$. Otherwise Eq. (B15) would have a non-zero DC term, in disagreement with Eq. (B3). Equation (B15) then reduces to

$$\begin{aligned}S_m(f)=\frac{1}{2}&|H(f-f_r)|^2|G(f-f_r)|^2\left[\lim_{T\to\infty}\frac{1}{T}\Lambda_T(0)\right.\\ &+\lim_{T\to\infty}\frac{1}{T}|\Lambda_T(2f_r)|\\ &\left.\cdot\cos\left(\pi+2\Phi_{opt}(f_r)-\Phi_{opt}(2f_r)\right)\right],\end{aligned} \tag{B16}$$

or

$$\begin{aligned}S_m(f)=\frac{1}{2}&|H(f-f_r)|^2|G(f-f_r)|^2\lambda_{avg}\\ &\cdot\left[1+\frac{|\tilde{P}_{opt}(2f_r)|}{\tilde{P}_{opt}(0)}\right.\\ &\left.\cdot\cos\left(\pi+2\Phi_{opt}(f_r)-\Phi_{opt}(2f_r)\right)\right].\end{aligned} \tag{B17}$$

Here we have used the fact that is $\Phi_\Lambda(f-f_r)$ is nonzero only when $f=0$, and $\Phi_\Lambda(-f_r)=-\Phi_{opt}(f_r)$. That is, the photoelectron spectral phase is equal to the spectral phase of the optical intensity profile. In accordance with Eq. (B3), the double-sided phase noise is the ratio of Eq. (B17) to the maximum DC offset (i.e., the second term in Eq. (B15) when its phase is set to zero):

$$S_\theta(f) = \frac{1}{2}\frac{\lambda_{avg}}{S_\lambda(f_r)}\left[1 + \frac{|\tilde{P}_{opt}(2f_r)|}{\tilde{P}_{opt}(0)}\right.$$
$$\left.\cdot \cos\left(\pi + 2\Phi_{opt}(f_r) - \Phi_{opt}(2f_r)\right)\right]. \quad (B18)$$

In terms of the photocurrent and microwave power, the single-sided phase noise (units rad$^2$/Hz) is given by

$$S_\theta(f) = \frac{2qI_{avg}|H_n(f_r)|^2 R}{P_\mu(f_r)}\left[1 - \frac{|\tilde{P}_{opt}(2f_r)|}{\tilde{P}_{opt}(0)}\right.$$
$$\left.\cdot \cos\left(2\Phi_{opt}(f_r) - \Phi_{opt}(2f_r)\right)\right]. \quad (B19)$$

where Eq. (10) has been used. The single-sideband phase noise (units dBc/Hz) is half this value, as given in Eq. (20). A similar expression may be derived for the amplitude noise, by letting $2\Phi_r + 2\Phi_H(f - f_r) + 2\Phi_G(f - f_r) + 2\Phi_\Lambda(f - f_r) = 0$ in Eq. (B15), and again normalizing by the maximum DC offset. In this case the minus sign in the brackets of Eq. (B19) becomes a plus sign.